\documentclass[fleqn,10pt]{wlscirep}

\usepackage{graphicx}
\usepackage{amsmath,amssymb,bm}
\usepackage[version=3]{mhchem} % Formula subscripts using \ce{}
\usepackage{subfigure}
\usepackage{siunitx}
\usepackage{dsfont}

\title{Generation of Subwavelength Plasmonic Nanovortices via Helically Corrugated Metallic Nanowires}

\author[1]{Changming Huang}
\author[1,*]{Xianfeng Chen}
\author[2,3]{Abiola O. Oladipo}
\author[2,$\sharp$]{Nicolae~C.~Panoiu}
\author[1,+]{Fangwei Ye}
\affil[1]{State Key Laboratory of Advanced Optical Communication Systems and Networks, Department of Physics and Astronomy, Shanghai Jiao Tong University, Shanghai 200240, China}

\affil[2]{Department of Electronic and Electrical Engineering,
University College London, Torrington Place, London WC1E7JE, UK}

\affil[3]{Bio-Nano Consulting, 338 Euston Road, NW1 3BT London, United Kingdom}

\affil[+]{fangweiye@sjtu.edu.cn} \affil[$\sharp$]{n.panoiu@ucl.ac.uk}
\affil[*]{xfchen@sjtu.edu.cn}

\begin{abstract}
We demonstrate that plasmonic helical gratings consisting of metallic nanowires imprinted with
helical grooves or ridges can be used efficiently to generate plasmonic vortices with radius much
smaller than the operating wavelength. In our proposed approach, these helical surface gratings
are designed so that plasmon modes with different azimuthal quantum numbers (topological charge)
are phase-matched, thus allowing one to generate optical plasmonic vortices with arbitrary
topological charge. The general principles for designing plasmonic helical gratings that
facilitate efficient generation of such plasmonic vortices are derived and their applicability to
the conversion of plasmonic vortices with zero angular momentum into plasmonic vortices with
arbitrary angular momentum is illustrated in several particular cases. Our analysis, based both on
the exact solutions for the electromagnetic field propagating in the helical plasmonic grating and
a coupled-mode theory, suggests that even in the presence of metal losses the fundamental mode
with topological charge $m=0$ can be converted to plasmon vortex modes with topological charge
$m=1$ and $m=2$ with a conversion efficiency as large as \SI{60}{\percent}. The plasmonic
nanovortices introduced in this study open new avenues for exciting applications of orbital
angular momentum in the nanoworld.
\end{abstract}
\begin{document}

\flushbottom
\maketitle
%\thispagestyle{empty}
%
%\noindent Please note: Abbreviations should be introduced at the first mention in the main text �C no abbreviations lists. Suggested structure of main text (not enforced) is provided below.

\section*{Introduction}

Optical vortices are light beams characterized by a phase change of an integer multiple of $2\pi$
along a closed path around the center of the beam, where the phase of the beam is undetermined
(phase singularity) and the field amplitude vanishes \cite{Desyatnikov2005291,Molina-Terriza,
Padgett}. The interest in such optical structures has dramatically increased since a connection
between the Laguerre-Gaussian laser modes and the orbital angular momentum (OAM) of light has been
established \cite{Allen}. In particular, it has been demonstrated that these beams and other
optical vortices carry an OAM of $m\hbar$ per photon \cite{Allen}, where $m$ is the so-called
topological charge of the optical vortex. This discovery has spurred intensive research interest
as in addition to its impact at a more fundamental science level, it has been realized that
OAM-carrying optical vortices could find a series of appealing applications to optical tweezers
\cite{tweezer1,tweezer2}, optical spectroscopy \cite{scopy}, digital imaging \cite{DSI}, and
quantum information processing \cite{quantum1,quantum2}. Importantly, significant advances in this
field have been facilitated by the fact that vortex beams can be readily generated by using a
multitude of experimental setups, including mode converters by astigmatic lenses \cite{mc1,mc2},
computer-synthesized holograms \cite{CGH, quantum2}, spiral-phase plates \cite{SPP}, angular
gratings \cite{grating}, and twisted elliptical \cite{kcs04s} and photonic crystal fibers
\cite{wkl12s,xwf14o}. One drawback of these methods, which severely hinders the extension of the
applications of optical vortices to the nanoscale, is that the diffraction limited propagation of
the vortical beams generated by these methods leads to spatially delocalized optical beams with
size significantly larger than the operating wavelength.

A recently introduced, promising approach towards optical beam engineering at subwavelength scale,
relies on plasmonic metasurfaces consisting of planar nanopatterned metallic structures. In
particular, thin metallic films in which nanoapertures with various shapes were milled in
\cite{prlnanoslit, olnanoslit,nlnanoslit,gorodetski2013generating} or planar arrays of metallic
nanoantennae \cite{metasurface} were used to experimentally generate plasmonic vortices. Despite
the fact that the phase of plasmonic vortices generated by these techniques can vary at
subwavelength scale, their size, however, was still much larger than the operating wavelength. As
an effective solution to this problem, in this report we demonstrate that one can generate
subwavelength optical vortices by first confining the optical field to subwavelength scale using a
metallic nanowire, the highly localized optical mode being then converted into an optical vortex
by means of a helical grating imprinted on the surface of the nanowire. Importantly, the
generation of subwavelength optical beams with zero angular momentum by using metallic nanowires
has been investigated both theoretically \cite{cnm10nl,wang2015transfer} and experimentally
\cite{wcs12acsn}, whereas the optical modes of helical gratings made of perfect conductors have
been studied in a recent theoretical work \cite{spoofvor}. Our theoretical and computational study
presented in this paper suggests that these ideas can be extended to the generation of
subwavelength optical vortices, namely one can employ plasmonic helical gratings to convert the
fundamental plasmonic mode of a uniform metallic nanowire to an optical beam carrying OAM, the
conversion efficiency being as large as \SI{60}{\percent} even in the presence of optical losses
in the metal.

\section*{Results}

\begin{figure}[ht]
\centering
\includegraphics[width=0.5\textwidth]{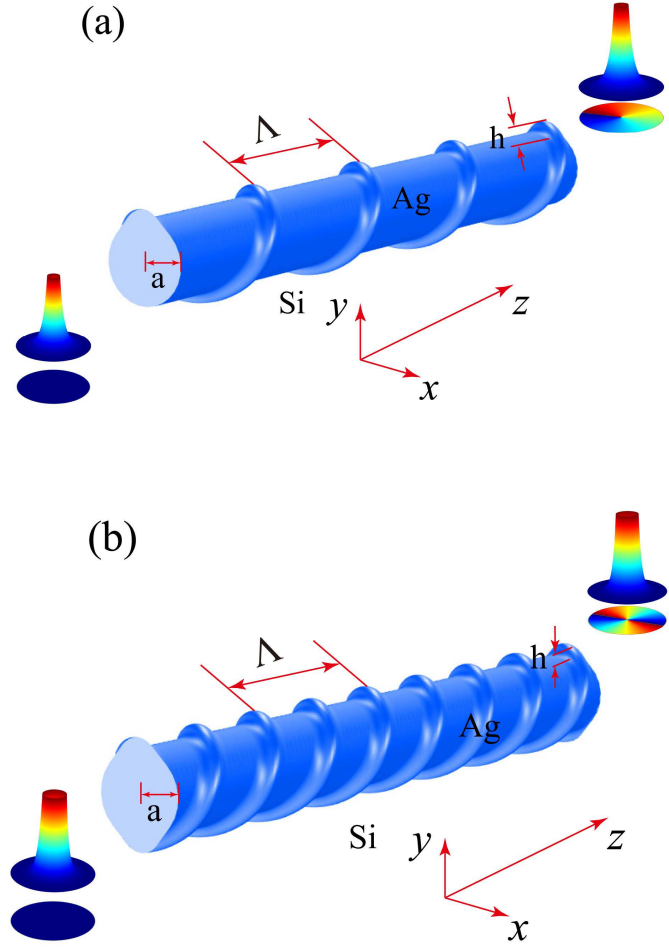}
\caption{\textbf{Schematic of a metallic (Ag) nanowire with a single and double helical surface
grating.} The helical surface gratings are designed to convert the fundamental mode with
topological charge $m=0$ to a plasmon vortex mode with topological charge $m=1$ (\textbf{a}) and
$m=2$ (\textbf{b}).} \label{fig:geometry}
\end{figure}

\textbf{Plasmonic helical grating.} The plasmonic helical grating designed to convert optical
modes of a uniform metallic nanowire is schematically depicted in Figure~\ref{fig:geometry}. It
consists of a metallic cylinder with constant radius, $a$, the surface of the cylinder being
engraved with a helical periodic grating with period, $\Lambda$, and height, $h\ll a$. Although it
is a challenging feat, the nanofabrication of such plamsonic helices has been recently reported in
several works \cite{Science09, PRL13}. In particular, rotating the ion beam of a focused-ion beam
system at high speed while periodically changing the radius of beam rotation \cite{FIB} or using
metal-assisted chemical etching \cite{chetch1,chetch2}, helical nanostructures similar to those
investigated in this work could be fabricated. We assume that the metallic nanowire is embedded in
a dielectric medium, the choice of materials considered in this paper being silver and silicon,
respectively. When a helical grating is imprinted on the nanowire, the fundamental plasmon mode,
which can be excited by an incident TM-polarized Gaussian beam, can be converted into vortex modes
provided that the wavevector mismatch between the fundamental mode and the desired vortical mode
is compensated by the properly designed grating.
%\begin{figure}[htbp]
%\centering
%\subfigure{\includegraphics[width=0.35\textwidth]{fig1.eps}}\hspace{0.0001\textwidth}
%\caption{Schematic of a metallic (Ag) nanowire with a single (a) and double (b) helical surface
%grating. The helical surface gratings are designed to convert the fundamental mode with
%topological charge $m=0$ to a plasmon vortex mode with topological charge $m=1$ (a) and $m=2$ (b).
%\label{fig:geometry}}
%\end{figure}

\noindent\textbf{Mode analysis of the uniform metallic nanowire.} The physical characteristics of
the mode conversion process depend on the properties of the optical modes of the uniform (constant
transverse section) metallic nanowire as well as the geometrical and electromagnetic properties of
the plasmonic helix. Regarding the optical modes of the nanowire, the main quantities that
determine the mode conversion efficiency are the field distribution and the modal propagation
constant. In the case of uniform metallic nanowires the optical modes can be readily obtained
analytically in cylindrical coordinates, $r$, $\phi$, and $z$ (see Supporting Information). For
convenience, we denote the optical modes as $|m\rangle e^{ik_0 \beta_m z}$, where $\beta_m$ is the
effective refractive index of the mode and $k_0=\omega/c$ is the wavenumber in vacuum at the
carrier frequency, $\omega$. The quantum number, $m=0,\pm1,\pm2,...$, also called topological
charge, defines the order of the mode and also describes the dependence of the optical field on
the azimuthal angle, via the exponential factor $e^{i m\phi}$.
\begin{figure}[htbp]
\centering \subfigure{\includegraphics[width=0.5\textwidth]{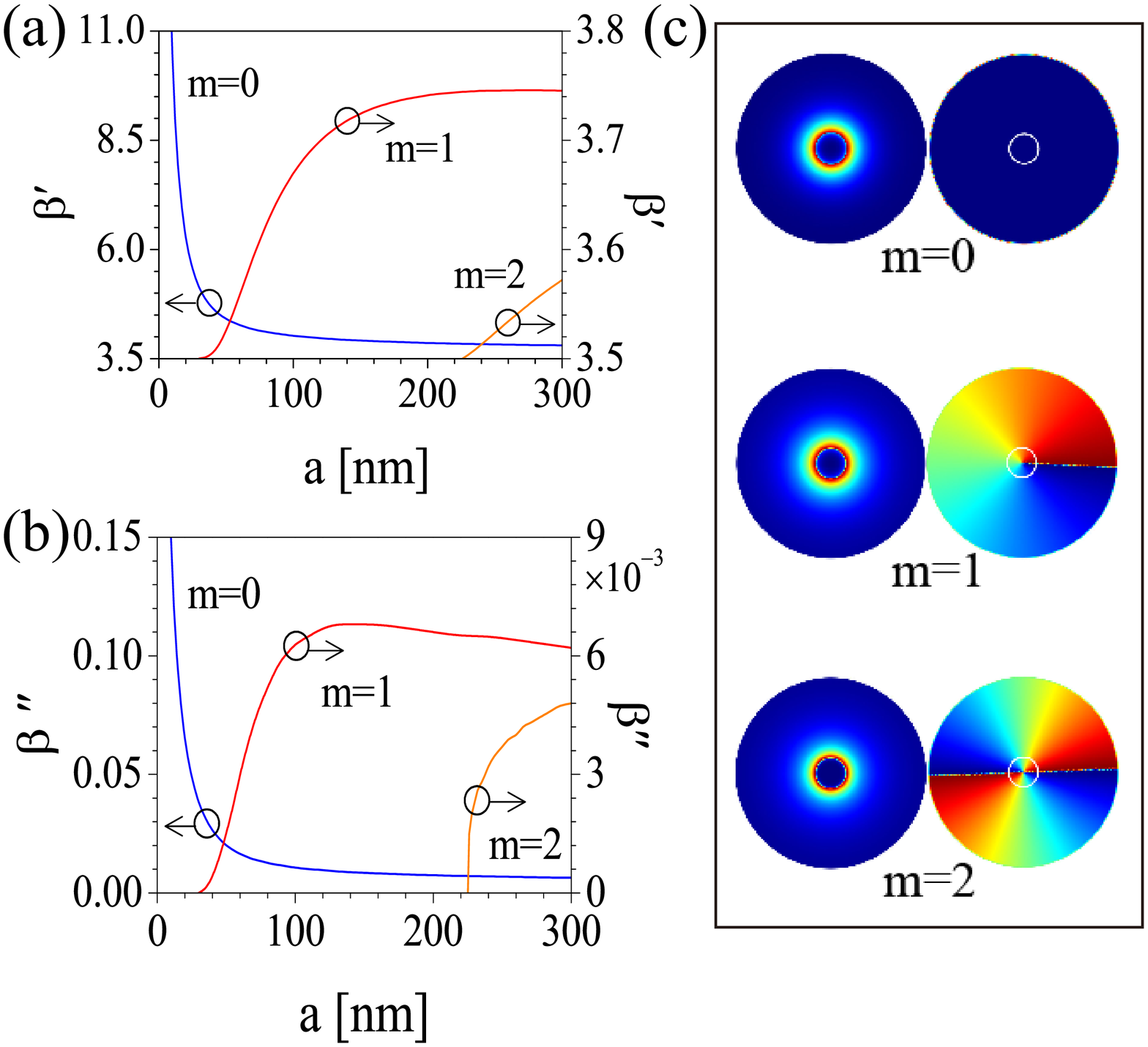}}\hspace{0.0001\textwidth}
\caption{\label{fig:modes}\textbf{Dispersion and field profiles of optical modes of metallic
nanowires.} (\textbf{a}), (\textbf{b}) Real and imaginary part of the effective mode index for the
first three modes, respectively. (\textbf{c}) Spatial profile of the amplitude and phase of the
modes with $m=0$, $m=1$, and $m=2$. In all panels, $\lambda=$~\SI{1500}{\nano\meter}.}
\end{figure}

\begin{figure}[t]
\centering
\includegraphics[width=0.35\textwidth]{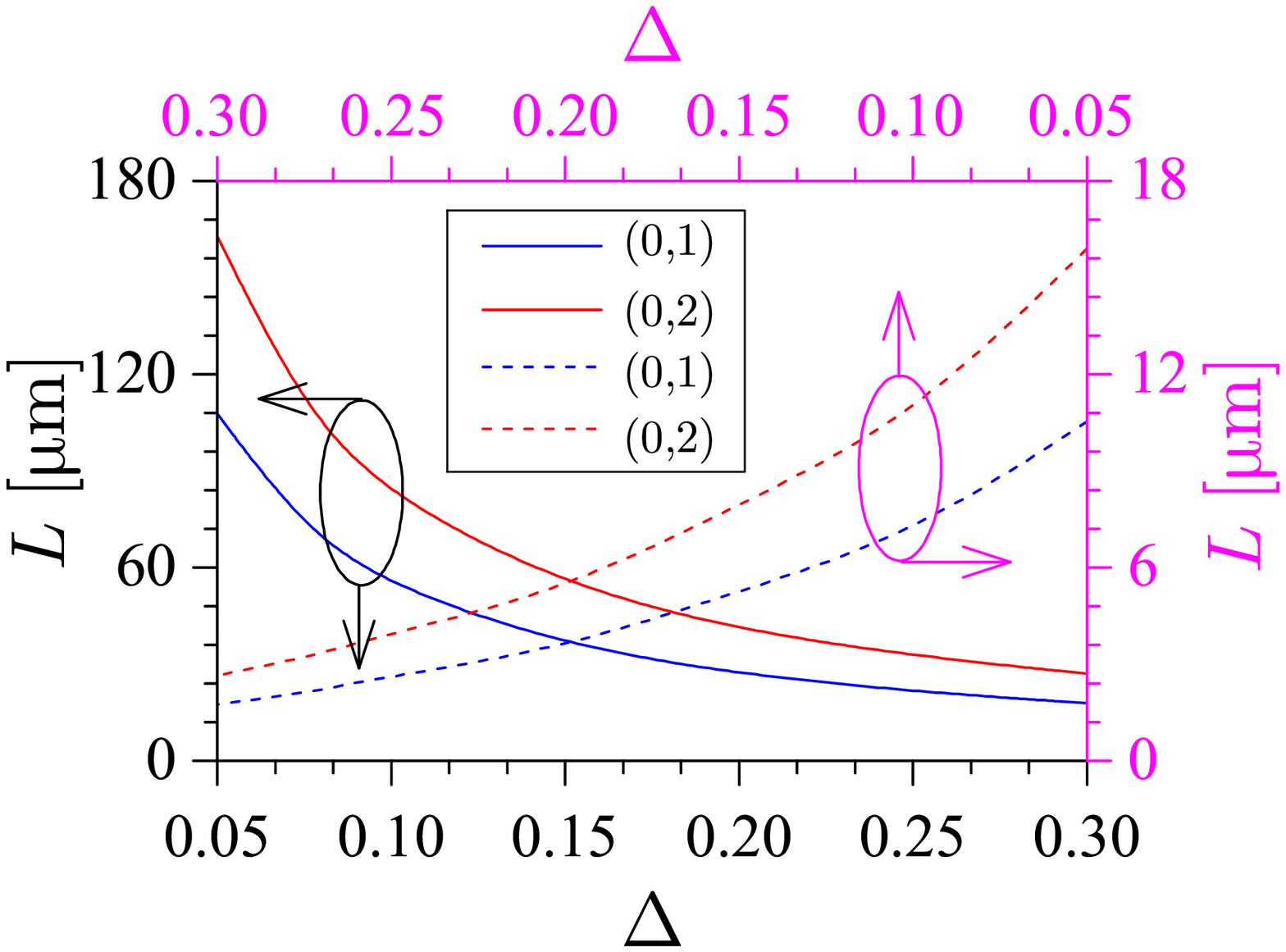}
\caption{\label{fig:conversion}\textbf{Coupled-mode theory analysis of mode interaction.} Coupling
length between the fundamental mode and the vortex modes with charge $m=1$ (blue lines) and $m=2$
(red lines). The grating is imprinted into the Si background (solid lines) and onto the Ag
nanowire (dashed lines). System parameters are $\lambda=$~\SI{1500}{\nano\meter} and
$h=$~\SI{20}{\nano\meter}. $\Lambda=$~\SI{5.03}{\micro\meter} and $a=$~\SI{110}{\nano\meter}
($\Lambda=$~\SI{5.09}{\micro\meter} and $a=$~\SI{250}{\nano\meter}) for the
$\vert0\rangle\rightarrow\vert1\rangle$ ($\vert0\rangle\rightarrow\vert2\rangle$) mode conversion
process.}
\end{figure}

The results of our mode analysis are presented in Figure~\ref{fig:modes}, where we plot the
variation of effective mode refractive index vs. the radius of the metallic nanowire, $a$, as well
as the modal amplitude and phase distributions, all determined for the first three modes, $m=0, 1,
2$. In this work we use the convention that the sign of the topological charge is positive
(negative) if the phase increases clockwise (counterclockwise) around the phase singularity.
Figure~\ref{fig:modes} shows that the fundamental mode does not have a cut-off frequency, namely
it exists for any radius of the nanowire, whereas for a given frequency vortical modes can only be
supported by a nanowire if its radius is larger than a certain critical value. Moreover, this
cut-off radius increases as the topological charge of the vortex mode increases. Note that the
cut-off radii of the nanowire corresponding to the vortices with charge $m=1$ and $m=2$ are still
well within the subwavelength regime, which suggests that metallic nanowires could potentially be
used to generate subwavelength optical vortices. Another important feature of the optical modes of
the nanowire is revealed by Figure~\ref{fig:modes}(b), namely their propagation loss dispersion.
Specifically, the propagation loss of the fundamental and vortical plasmon modes, which is
proportional to $\beta_{m}^{\prime\prime}$, has contrasting dependence on the nanowire radius:
while in the case of the fundamental mode the propagation loss decreases sharply with the radius,
in the case of vortical modes there is a steep increase with the radius near the mode cut-off,
which is followed by a region in which the propagation loss decreases slowly with the radius.
Here, $z^{\prime}$ ($z^{\prime\prime}$) represents the real (imaginary) part of the complex number
$z$. In particular, at $\lambda=$~\SI{1500}{\nano\meter} the figure of merit of plasmon modes,
defined as $\beta_{m}^{\prime\prime}/\beta_{m}^{\prime}$, is \num{\sim e-2} and \num{\sim e-3} for
the fundamental mode and vortices, respectively.

\noindent\textbf{Coupled-mode theory.} The phase-matching condition for efficient mode conversion
can be readily derived by using a vectorial coupled-mode theory (CMT). In the standard framework
of CMT, the total electromagnetic field in the perturbed nanowire (helical grating) is expressed
as linear superposition of the modes of the uniform nanowire,
$\mathbf{E}(\mathbf{r})=\sum_{m}a_{m}(z)\mathbf{e}_{m}(r)e^{i(k_{0}\beta_m z+m\phi-\omega t)}$ and
$\mathbf{H}(\mathbf{r})=\sum_{m} a_{m}(z)\mathbf{h}_{m}(r)e^{i(k_{0}\beta_m z+m\phi-\omega t)}$,
where $\mathbf{e}_{m}(r)$ and $\mathbf{h}_{m}(r)$ are the electric and magnetic fields of the mode
with topological charge, $m$, of the unperturbed nanowire, respectively. The main result provided
by the CMT is the coupled-mode equations (CMEs), which govern the dependence of the mode
amplitudes, $a_m(z)$, on the propagation distance (see Supporting Information for the derivation
of these equations):
\begin{equation}
\label{eq:cme} \frac{d a_m(z)}{dz}=i{\displaystyle\sum_{p}}K_{mp}a_p(z)e^{ik_{0}\Delta
\beta_{mp}z} e^{i\varrho_{mp}\phi},
\end{equation}
where $\Delta \beta_{mp}=\beta_p-\beta_m-\sigma\kappa$ and $\varrho_{mp}=p-m-\sigma$ quantify the
phase mismatch between modes $|m\rangle$ and $|p\rangle$, in the longitudinal and transverse
directions, respectively, the reduced wave vector, $\kappa=2\pi/k_{0}\Lambda$, and $\sigma$ is the
order of the helix, namely the order of the symmetry point group $\bm{\mathcal{C}}_{n}$ of the
helix (e.g., $\sigma=1$ and $\sigma=2$ for a single- and a double-helix, respectively). The
coefficients $K_{mp}$ describe the coupling between the two interacting modes, and is given by the
following overlap integral:
\begin{equation}
\small{
\label{eq:cc} K_{mp}=\frac{\omega}{4\sqrt{P_p P_m}} \int{\delta\epsilon(\mathbf{r})
\left[\mathbf{e}_{p,\perp} \cdot \mathbf{e}_{m,\perp}
-\frac{\epsilon{(\mathbf{r}_{\perp})}}{\tilde{\epsilon}(\mathbf{r})} \mathbf{e}_{p,z}\cdot
\mathbf{e}_{m,z}\right]}d\mathbf{r}_{\perp},
}
\end{equation}
In this equation $P_{i}$ is the normalization power of mode $\vert i\rangle$, the dielectric
perturbation $\delta\epsilon(\mathbf{r})$ is the difference between the dielectric constant of the
uniform nanowire and helical grating, $\epsilon(\mathbf{r}_{\perp})$ and
$\tilde{\epsilon}(\mathbf{r})$ are the dielectric constant of the uniform nanowire and helical
grating, respectively, and the symbol ``$\perp$'' indicates the transverse component of a vector.

It can be clearly seen from Eq.~(\ref{eq:cme}) that in order to achieve an efficient energy transfer
between the modes $\vert m\rangle$ and $\vert p\rangle$, their longitudinal and transverse
phase-mismatch must be simultaneously compensated by a proper choice of the helical perturbation,
$\delta\varepsilon(\mathbf{r})$. To be more specific, we assume that the dielectric constant of
the plasmonic system can be expressed as follows:
\begin{eqnarray}
\label{eq:grating}
\tilde{\varepsilon}(r,\phi,z)=\varepsilon(r,\phi)+\delta\varepsilon(r,\phi,z) %\nonumber\\
=\left\{
\begin{array}{cllc}
&\varepsilon_m, & {r<a},\\
&\varepsilon_d[1+\Delta \sin\sigma(\phi-\kappa z)], & {a \leq r \leq a+h},\\
&\varepsilon_d, & {r>a+h.}
\end{array}
\right.
\end{eqnarray}
The parameters $\Delta$ and $\kappa$ represent the grating perturbation strength and the grating
wave vector, respectively. Specifically, Eq.~(\ref{eq:grating}) describes a helix imprinted onto the
background region (chosen to be silicon in our case, $\varepsilon_d=12.25$) within a cylindrical
shell with thickness, $h$. Similarly, the helix can be imprinted into the metallic region (silver
in our case, $\varepsilon_m=-125$ at $\lambda=$~\SI{1500}{\nano\meter}) as well. We have
investigated both cases, the main conclusions being qualitatively similar.

\begin{figure}[t]
\centering \subfigure{\includegraphics[width=0.45\textwidth]{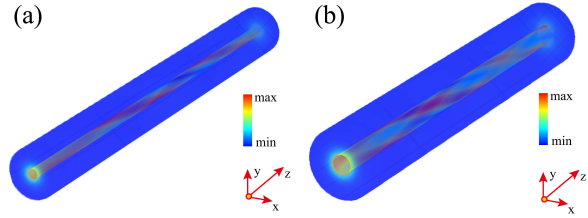}}\hspace{0.0001\textwidth}
\caption{\label{fig:field}\textbf{Field distributions obtained by solving the 3D Maxwell equations.} The
numerical simulations were performed for (\textbf{a}) a single- and (\textbf{b}) double-helix structure. The system
parameters in the left (right) panel are $\Lambda=$~\SI{5.03}{\micro\meter} and
$a=$~\SI{110}{\nano\meter} ($\Lambda=$~\SI{5.09}{\micro\meter} and $a=$~\SI{250}{\nano\meter}). In both
cases $\lambda=$~\SI{1500}{\nano\meter} and $h=$~\SI{20}{\nano\meter}.}
\end{figure}

\begin{figure*}[t]
\centering \subfigure{\includegraphics[width=0.8\textwidth]{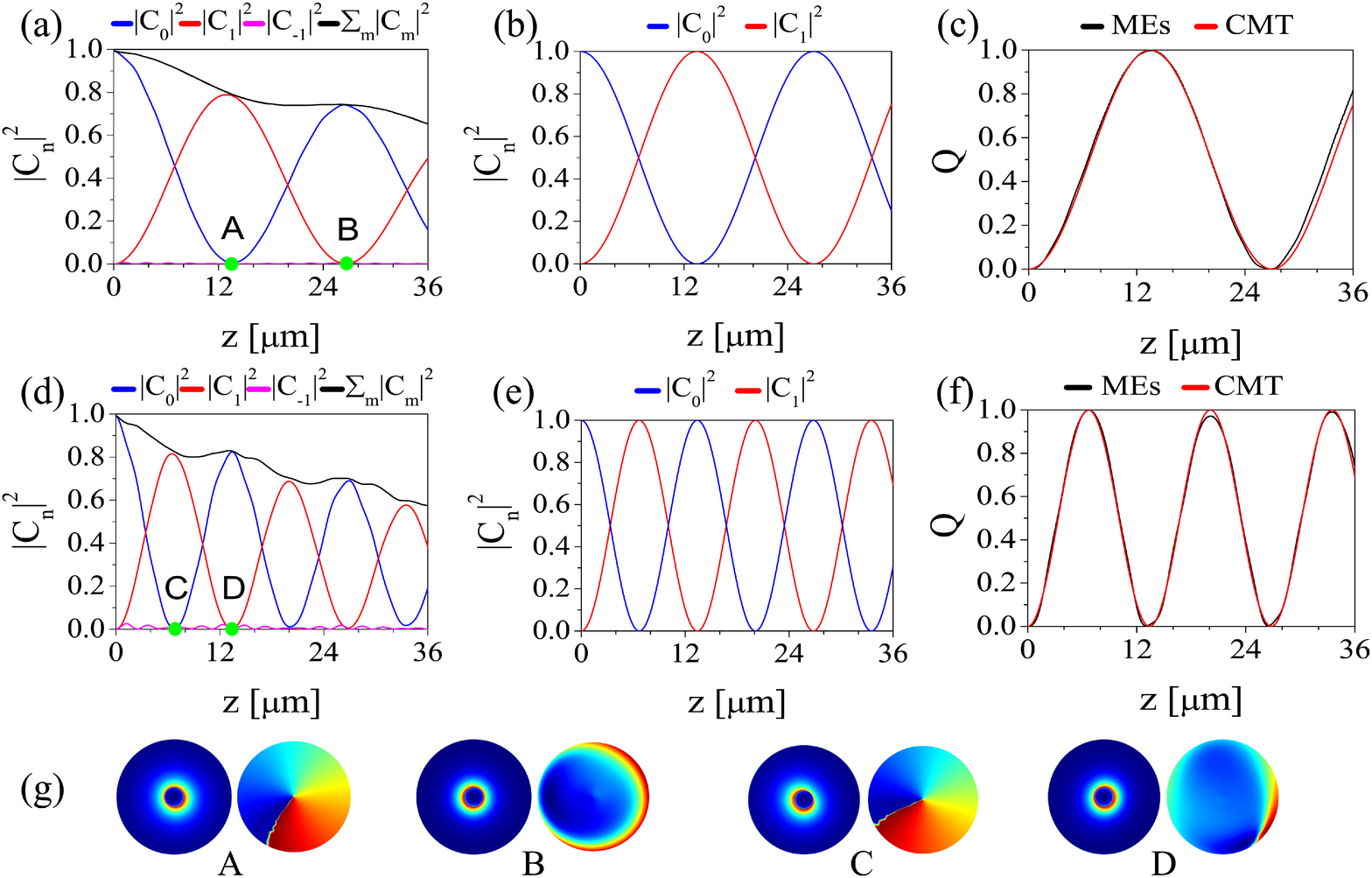}}\hspace{0.0001\textwidth}
\caption{\label{fig:m1}\textbf{Generation of nanovortices with topological charge} $m=1$.
Variation of mode weight, $|C_n|^2$, and effective topological charge, $Q$, vs. propagation
distance, $z$, calculated for grating strength $\Delta=0.1$ (top panels) and $\Delta=0.2$ (middle
panels). Results in panels (a) and (d) are found by solving the 3D Maxwell equations whereas those
in panels (b) and (e) are calculated using the CMT. (g), The amplitude (left panels) and phase
(right panels) structure of the plasmonic field, calculated at four distances, as shown in panels
(a) and (d). Other parameters are: $\lambda=$~\SI{1500}{\nano\meter}, $h=$~\SI{20}{\nano\meter},
$\Lambda=$~\SI{5.03}{\micro\meter}, and $a=$~\SI{110}{\nano\meter}.}
\end{figure*}

As explained above, efficient mode conversion occurs provided that the grating wave vector,
$\kappa$, and the helix order, $\sigma$, are chosen so as both the longitudinal and transverse
phase difference between the two interacting modes are compensated. For example, in order to
convert the fundamental mode ($m=0$) into a vortex mode with charge $m$, the pitch of the helix
must satisfy the relation $\Lambda=\frac{\displaystyle 2\pi\sigma}{\displaystyle
k_0\vert\beta_0-\beta_m\vert}$ and the order of the helix should be $\sigma=m$. Importantly, the
handedness of the helical grating determines the sign of the generated vortex. This can be easily
understood by expanding the perturbation as $\delta \varepsilon(\mathbf{r})=\varepsilon_d \Delta
\sin\sigma(\phi-\kappa z)=\varepsilon_d \Delta [e^{i\sigma(\phi-\kappa z)}-e^{-i\sigma(\phi-\kappa
z)}]/2i$. As the nanowire excitation mode is $\vert0\rangle e^{i k_0 \beta_0 z}$, the only two
vortices that could be excited via the helical grating described by Eq.~(\ref{eq:grating}) are
$\vert\sigma\rangle e^{i k_0(\beta_0-\sigma\kappa)z}$ and $\vert-\sigma\rangle
e^{ik_0(\beta_0+\sigma\kappa)z}$. However, the latter vortex, which has a negative charge, has a
larger effective mode index as compared to that of the fundamental mode and therefore it is not
phase-matched to the excitation mode. Therefore, it is expected that in this case only the vortex
with positive charge, $m=\sigma$, is generated. On the other hand, if the handedness of the helix
is reversed, a negatively charged vortex would be generated.

A key quantity that characterizes the mode conversion process is the conversion length, $L$, which
is defined as the distance over which the fundamental mode is completely converted into a vortex
mode $\vert m\rangle$. It can be easily shown that Eq.~(\ref{eq:cme}) implies that the conversion
length is given by $L=2\pi/|K_{0m}|$, which suggests that a larger optical coupling between the
interacting modes should lead to a shorter conversion length. This conclusion is fully supported
by the results presented in Figure~\ref{fig:conversion}, where we plot the dependence of the
coupling length on the grating perturbation strength, $\Delta$, for the mode conversion processes
$\vert0\rangle\rightarrow\vert1\rangle$ and $\vert0\rangle\rightarrow\vert2\rangle$. This figure
shows that the coupling length decreases when the grating strength, $\Delta$, increases, which
suggests that it could be possible to reach an operation regime in which the coupling length is
smaller than the characteristic loss length of the interacting modes by simply increasing the
grating strength. Moreover, it can be seen that the coupling length is larger when the grating is
imprinted into the silicon background as compared to the case when it is engraved onto the
metallic nanowire. This is an expected result as in the latter case there is a larger perturbation
of the dielectric constant of the system, due to the fact that the dielectric constant of silver
is significantly larger than that of silicon. In addition, for both types of helical gratings, the
coupling length increases with the charge of the generated vortex, chiefly due to the fact that
the overlap between the fundamental mode and the vortex modes increases with the topological
charge.
\begin{figure*}[t]
\centering \subfigure{\includegraphics[width=0.8\textwidth]{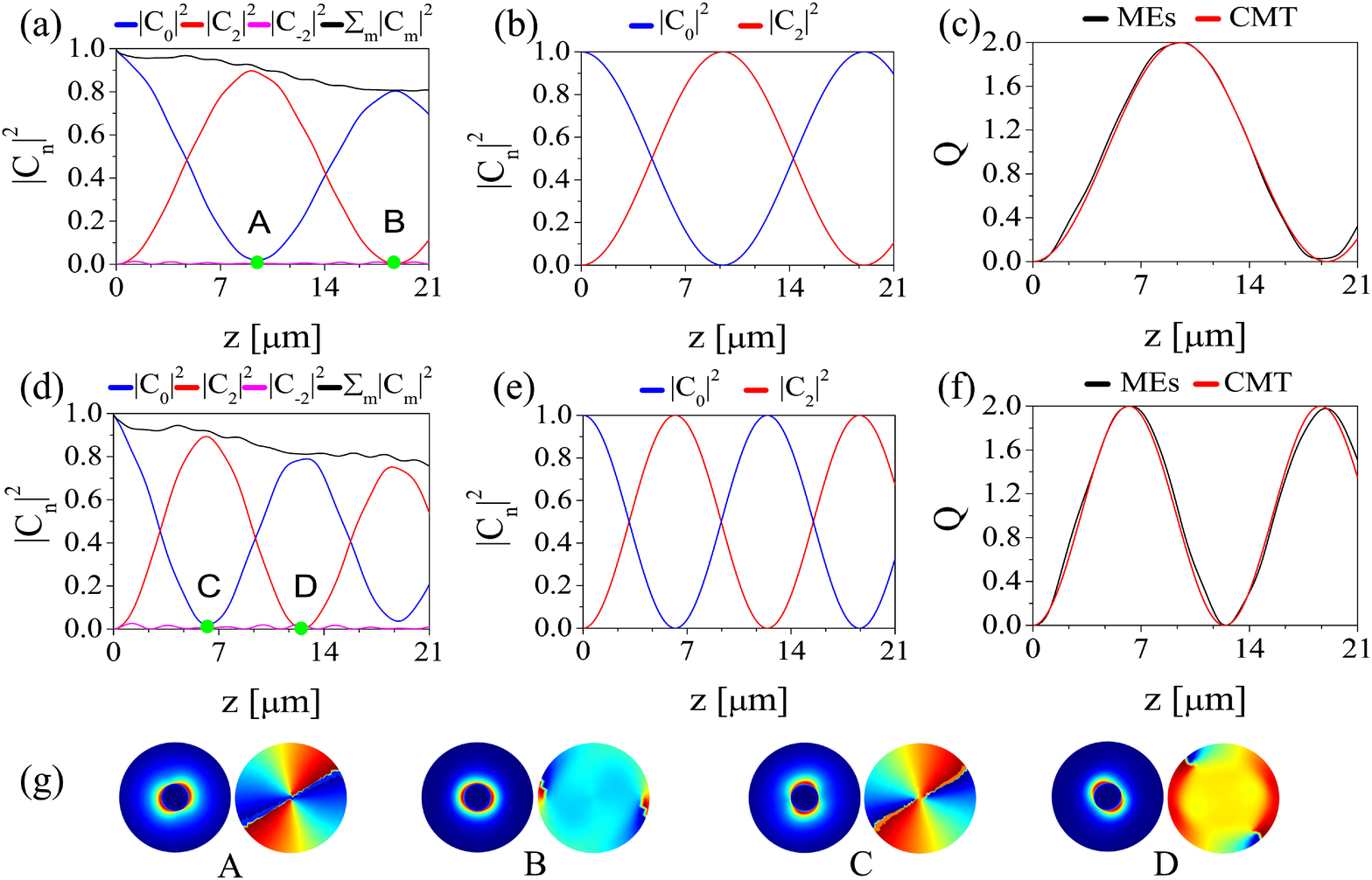}}\hspace{0.0001\textwidth}
\caption{\label{fig:m2}\textbf{Generation of nanovortices with topological charge} $m=2$.
Variation of mode weight, $|C_n|^2$, and effective topological charge, $Q$, vs. propagation
distance, $z$, calculated for grating strength $\Delta=0.2$ (top panels) and $\Delta=0.3$ (middle
panels). Results in panels (a) and (d) are found by solving the 3D Maxwell equations whereas those
in panels (b) and (e) are calculated using the CMT. (g), The amplitude (left panels) and phase
(right panels) structure of the plasmonic field, calculated at four distances, as shown in panels
(a) and (d). Other parameters are: $\lambda=$~\SI{1500}{\nano\meter}, $h=$~\SI{20}{\nano\meter},
$\Lambda=$~\SI{5.09}{\micro\meter}, and $a=$~\SI{250}{\nano\meter}.}
\end{figure*}

\noindent\textbf{Rigorous numerical simulations.} Encouraged by these results derived from the
CMT, we sought to validate them by using rigorous electromagnetic numerical simulations based on
the fully 3D exact solutions of the Maxwell equations. To this end, we determined first from the
CMT the pitch, $\Lambda$, of the helical grating by imposing the condition that the two
interacting optical modes are phase-matched. Then, we launched the fundamental mode ($m=0$) at the
input facet of a plasmonic helical grating designed to phase-match this mode and a specific
vortical mode ($m\neq0$), the total 3D field distribution being determined by numerically solving
the Maxwell equations \cite{COMSOL}. As an alternative to this rigorous approach, we used
Eq.~(\ref{eq:cme}) to calculate the amplitudes of the interacting modes and, subsequently, the 3D
field distribution. As a result, this computational analysis would provide valuable insights into
the regime in which the predictions of the CMT are valid.

We have followed the computational approach just described and studied the vortex generation
processes $\vert0\rangle\rightarrow\vert1\rangle$ and $\vert0\rangle\rightarrow\vert2\rangle$, the
corresponding field dynamics being presented in Figures~\ref{fig:field}(a) and \ref{fig:field}(b),
respectively. These field profiles clearly illustrate the formation of optical vortices. In
addition, they show an important feature of the mode conversion process, namely, unlike the case
of helical optical fibers, the length over which the fundamental mode is converted into a vortex
mode is larger but comparable to the pitch of the helix, $\Lambda$.

\noindent\textbf{Mode conversions.}
Let us now analyze more closely the $\vert0\rangle\rightarrow\vert1\rangle$ conversion process.
The physical quantities most suitable for characterizing the efficiency of this process are the
effective topological charge of the field, $Q=\omega L_{z}/U_{z}$, where the total orbital angular
momentum of the mode, $L_{z}$, and its intensity, $U_{z}$, are given by
\begin{align}
L_z&=\frac{\varepsilon_0}{2}\int \mathbf{r}\times(\mathbf{E}^*\times\mathbf{B})\cdot \hat{\bm{\mathsf{e}}}_z d\mathbf{r}_{\perp}, \\
U_z&=\frac{\varepsilon_0 c}{2\beta}\int (\mathbf{E}^*\times\mathbf{B})\cdot
\hat{\bm{\mathsf{e}}}_z d\mathbf{r}_{\perp},
\end{align}
and the modal weight, $C_{m}$, defined as
\begin{equation}
C_m=\frac{\displaystyle\int (\mathbf{E} \times \mathbf{H}_{m}^*)\cdot \hat{\bm{\mathsf{e}}}_z
d\mathbf{r}_{\perp}}{\displaystyle\int (\mathbf{E}_m \times \mathbf{H}_m^*)\cdot
\hat{\bm{\mathsf{e}}}_z d\mathbf{r}_{\perp}},
\end{equation}
which quantifies the relative amount of power flowing in the mode, $m$. The dependence of these
physical quantities on the propagation distance, $z$, is presented in Figure~\ref{fig:m1}. We
considered two helical gratings with values of the grating strength, namely $\Delta=0.1$ (top
panels) and $\Delta=0.2$ (middle panels). As the figure shows, upon the propagation of the
fundamental mode in the grating, its weight, $|C_0|$, gradually decreases with $z$, whereas the
weight of the $\vert1\rangle$ vortex, $|C_1|$, increases until the maximum conversion is reached
at a quarter of coupling length, $z=L/4$. The maximum mode conversion corresponds to the point
\textit{A} (\textit{C}) in Figure~\ref{fig:m1}(a) [Figure~\ref{fig:m1}(d)]. The corresponding
intensity and phase distribution at these points are shown in Figure~\ref{fig:m1}(g). Because the
mode interaction increases with the grating strength, the coupling length should decrease with
$\Delta$ [compare the location of points \textit{A} and \textit{C} in Figure~\ref{fig:m1}(a) and
Figure~\ref{fig:m1}(d), respectively]. This is the expected dynamics of the plasmonic field
indeed, as the grating was designed to phase match the interaction of the $\vert0\rangle$ and
$\vert1\rangle$ modes. Beyond the maximum mode conversion point the power flow between the two
modes is reversed. Note, however, one interesting idea revealed by Figure~\ref{fig:m1}(d): the sum
of mode powers weakly increases whenever the transformation of the vortex mode into the
fundamental mode occurs, indicating that some energy of the radiation modes is fed back into the
nanostructure during the back-conversion.

During the mode conversion, as expected, the effective topological charge of the total field, $Q$,
displays the same evolution as $|C_1|$, approaching almost unit value at the coupling length
[Figures~\ref{fig:m1}(c) and \ref{fig:m1}(f)]. Note that, due to its phase-mismatched with the
fundamental or other modes, the mode $\vert-1\rangle$, namely the vortex with charge $m=-1$, is
not excited. It should be noted that, however, due to the excitation of radiative modes and
possibly other vortical modes with larger topological charge during the mode conversion, not all
of the energy of the fundamental mode is transferred to the $\vert1\rangle$ vortex, thus both
$|C_1|^2<1$ and $Q<1$ at $z=L$. Despite this, a remarkably large conversion efficiency can be
achieved, more than \SI{80}{\percent} of the energy of the fundamental mode being transferred to
the $\vert1\rangle$ vortex. We also studied the plasmonic field evolution and the corresponding
physical quantities that characterize its dynamics by using the CMT, the results being presented
in Figures~\ref{fig:m1}(b) and \ref{fig:m1}(e). One can see that the predictions of the CMT
regarding the coupling length are in very good agreement with the results of direct simulations.
As expected, the conversion efficiency calculated using the CMT agrees less with the simulation
results, primarily because we included in the CMT calculations only the two interacting modes.

Our analysis shows that a single-helical metallic nanowire with length $L$ can be viewed as a
source of unit-charge nanovortices, the radius of the generated vortices being roughly equal to
the radius of the nanowire. In the case of Figure~\ref{fig:m1}, vortices have a radius of about
\SI{110}{\nano\meter}, which is more than an order of magnitude smaller than the operating
wavelength, $\lambda=$~\SI{1500}{\nano\meter}. Similarly, metallic nanowires with double-helix
surface corrugation, that is $\sigma=2$ in Eq.~(\ref{eq:grating}), could be used to generate optical
vortices with topological charge equal to $2$. This is clearly demonstrated by the plots presented
in Figure~\ref{fig:m2}, which summarize the results of our analysis of the field dynamics in a
double-helix plasmonic grating designed to phase-match the fundamental mode and the
$\vert2\rangle$ vortex. However, it should be mentioned that, the higher the order of the desired
vortex is, the larger its cut-off value for a specific radius will be. This suggests that the size
of the generated vortices increases with the order of the vortex. Despite this, the generated
doubly-charged vortex shown in Figure~\ref{fig:m2} has a radius of $250~\textrm{nm}$, which is
still significantly smaller than the wavelength.
\begin{figure}[htbp]
\centering \subfigure{\includegraphics[width=0.33\textwidth]{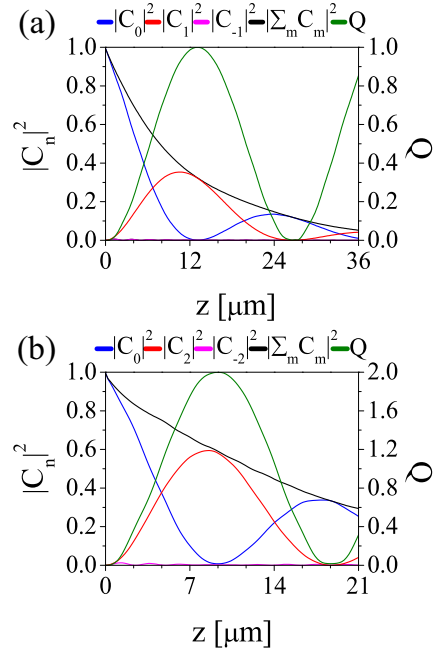}}\hspace{0.0001\textwidth}
\caption{\label{fig:loss}\textbf{Generation of optical vortices with charge} $m=1$ (\textbf{a})
\textbf{and} $m=2$ (b) \textbf{in lossy metallic helical gratings.} The left (right) panel
corresponds to the case presented in the top panels of Figure~\ref{fig:m1} (Figure~\ref{fig:m2}).}
\end{figure}

Finally, we considered the influence of the metal losses on the generation of nanovortices.
Metallic losses can be particularly detrimental to the vortex generation process as the
electromagnetic energy could be dissipated over an effective loss length that is smaller than the
coupling length, so that no vortices would be generated. Fortunately, one can decrease the
coupling length by simply increasing the grating strength, $\Delta$, or by increasing its height,
$h$ [see Figure~\ref{fig:conversion}; also compare the coupling lengths in Figures~\ref{fig:m1}(a)
and \ref{fig:m1}(d) or those in Figures~\ref{fig:m2}(a) and \ref{fig:m2}(d)], so that one can
easily achieve significant mode conversion before the fundamental mode decays. The conversion of
the fundamental mode into the $\vert1\rangle$ and $\vert2\rangle$ nanovortices, when metallic
losses are fully incorporated in simulations, are shown in Figures~\ref{fig:loss}(a) and
\ref{fig:loss}(b), respectively. This figure suggests that a surprisingly large mode conversion
efficiency can be achieved, namely $\sim$\SI{35.3}{\percent} and $\sim$\SI{60.2}{\percent} for the
vortices with topological charge $m=1$ and $m=2$, respectively. The fact that lower efficiency is
achieved in the case of the vortex with $m=1$ can be explained by noticing that the optical field
of this vortex is more confined around the metallic naowire and therefore the corresponding
optical losses are larger.

\section*{Conclusions}
In conclusion, in this study we have introduced a new type of sources of nanovortices, namely,
metallic cylinders with deep-subwavelength radius and helically corrugated surfaces. With a proper
selection of the period of the helix, these helical gratings can be used to generate nanoscale
vortices with various topological charge. A coupled-mode theory of mode conversion was developed,
its predictions being in excellent agreement with the conclusions of direct simulations based on
the full set of Maxwell equations. The plasmonic nanovortices introduced in this study might
extend a series of appealing applications of OAM-carrying light beams to the nanoworld, such as nanoscaled
optical spanners \cite{tweezer1} and digital imagining \cite{DSI}, as well as the integrated quantum information processing \cite{quantum1}.

%\bibliography{heliextoSR}
%\bibliographystyle{plain}

\section*{Acknowledgements}

The work of C. Huang and F. Ye was supported by the NSFC, Grant No. 11104181 and 61475101.
A.~O.~Oladipo and N.~C.~Panoiu wish to acknowledge support from the EU-FP7 (NMP-2011-280516,
"VSMMART-Nano"). X. Chen appreciates the support of NSFC, Grant No. 61125503.

\section*{Correspondence}
Correspondence and requests for materials should be addressed to Fangwei Ye (email:
fangweiye@sjtu.edu.cn), Xianfeng Chen (email: xfchen@sjtu.edu.cn), or Nicolae C. Panoiu (email:
n.panoiu@ucl.ac.uk).

\section*{Author contributions statement}
F. Y. and N. C. P. conceived the idea of the paper; F. Y., C. H. and A. O. O. performed the
numerical simulations and prepared the data. X. C. supervised the numerical simulations work. F.
Y. and N. C. P. wrote the paper and all authors reviewed the manuscript.

\section*{Additional information}

\noindent\textbf{Supplementary information} accompanies this paper on http://www.nature.com/scientificreport.

\noindent\textbf{Competing financial interests:} The authors declare no competing financial interests.

\end{document}